# Influence of the sodium/proton replacement on the structural, morphological and photocatalytic properties of titanate nanotubes

V. Bem[a], M.C. Neves[b], M.R. Nunes[c], A.J. Silvestre[a] and O.C. Monteiro[d*]

[a]Instituto Superior de Engenharia de Lisboa, Department of Physics and ICEMS, R. Conselheiro Emídio Navarro 1, 1959-007 Lisbon, Portugal

[b]University of Aveiro, Department of Chemistry, CICECO, 3810 – 193 Aveiro, Portugal

[c]University of Lisbon, Faculty of Sciences, Department of Chemistry and Biochemistry, CCMM, Campo Grande, 1749 – 016 Lisbon, Portugal

[d]University of Lisbon, Faculty of Sciences, Department of Chemistry and Biochemistry, CQB, Campo Grande, 1749 – 016 Lisbon, Portugal

**Abstract**

Titanate nanotubes (TNT) with different sodium contents have been synthesised using a hydrothermal approach and a swift and highly controllable post-washing processes. The influence of the sodium/proton replacement on the structural and morphological characteristics of the prepared materials was analysed. Different optical behaviour was observed depending on the $Na^+/H^+$ samples' content. A band gap energy of 3.27±0.03 eV was estimated for the material with higher sodium content while a value of 2.81±0.02 eV was inferred for the most protonated material, which therefore exhibits an absorption edge in the near visible region. The point of zero charge of the materials was determined and the influence of the sodium content on the adsorption of both cationic and anionic organic dyes was studied. The photocatalytic performance of the TNT samples was evaluated in the rhodamine 6G degradation process. Best photodegradation results were obtained when using the most protonated material as catalyst, although this material has shown the lowest R6G adsorption capability.

**Keywords**: Titanate nanotubes; sodium/proton exchange; band gap energy; photocatalysis; rhodamine 6G

[*] Corresponding author: Phone:+351 217500865; Fax:+351 217500088; E-mail: ocmonteiro@fc.ul.pt

# 1 Introduction

Semiconducting $TiO_2$ has been thoroughly investigated as photocatalyst for the treatment of industrial wastewaters and polluted air [1-3]. Its major drawback is however its high charge recombination rate and its wide band gap (*anatase*, *ca.* 3.2 eV), which limits the electron and hole photogeneration under visible irradiation. Therefore, the synthesis of $TiO_2$-*based* materials with either a broader range of light absorption and/or a lower charge recombination rate would be an important achievement toward the development of successful photoactive materials.

Inspired by the discovery of carbon nanotubes, one-dimensional nanostructured materials have become a research topic owing to their unusual properties and potential applications. Among the various semiconductors, titanate nanotubes (TNTs) have been subject of interest because of their cheap fabrication, unique one-dimensional nanostructure, high surface area and electrical conductivity [4,5]. Moreover, TNT based materials are able to combine the properties and applications of conventional $TiO_2$ nanoparticles (*e.g.*, photocatalytic activity) with the properties of layered titanates (*e.g.*, ion-exchange facility). As result, they have the potential to be used in a wide variety of applications, including photocatalysis [6], as substrates decorated with different active catalysts [7], dye-sensitized solar cells [8] transparent optical devices [9], and gas or humidity sensors [10]. Particularly important could be their potential use in photocatalytic applications since they have a similar structure to titanium dioxide but larger surface area. Additionally, some studies suggest that TNT, because of their elongated morphology, exhibit improved photogenerated charged carrier separation than $TiO_2$ nanoparticles [4, 5].

Since the discovery of the $TiO_2$-*based* alkaline hydrothermal chemical route, by Kasuga et al. [11], the mechanism of TNT formation has been the subject of intense research [12-14]. Nanostructured tubular titanates are characterized by an open mesoporous morphology, high specific surface area and good ion-exchange properties. The walls of the titanate nanotubes have a characteristic multilayered structure consisting of edge- and corner-sharing $TiO_6$ octahedra building up zigzag structures with the sodium ions located between the $TiO_6$ layers. Protonated nanostructured titanates can be obtained by a posterior sodium/proton exchange process. The detailed crystal structure of TNT is still unclear, however the generalized formula of the sodium titanate nantubes is assumed to be $Na_2Ti_3O_7$ and that of the protonated titanates $H_2Ti_nO_{(2n+1)} \cdot xH_2O$ [15].

Despite the success of the alkaline hydrothermal chemical method in the TNT synthesis, the use of either distinct commercial or home synthesised sources of titania powders as precursor materials frequently leads to nanostructures with different microstructural features, rendering the reproducibility of the TNT synthesis process highly dependent on the raw crystalline $TiO_2$ starting material and, therefore, far from being well established.

In this work a swift synthesis approach is proposed to prepare homogenous titanate nanostructures with different sodium contents. The chemical route used is highly reproducible and avoids the use of titania as precursor material. The TNT sodium content, and the consequent variation of the material properties, was easily achieved through control of the washing parameters, mainly the *pH* of the filtrate solution. The influence of the sodium/proton content on the TNT structure, morphology and other physical and chemical properties was analysed. The potential applicability of these materials for organic pollutants remediation processes was investigated, by studying their adsorption ability and photocatalytic performance in the rhodamine 6G (R6G) degradation reaction.

## 2. Materials and Methods

All reagents were of analytical grade (Aldrich and Fluka) and were used as received. The solutions were prepared with Millipore Milli-Q ultra pure water.

### 2.1. Materials

#### 2.1.1. TNT precursor synthesis

The TNT precursor was prepared using a procedure previously reported [16]. A titanium trichloride solution (10 wt.% in 20-30 wt.% HCl) diluted in a ratio of 1:2 in standard HCl solution (37 %) was used as titanium source. To this solution a 4 M ammonia aqueous solution was added dropwise under vigorous stirring, until complete precipitation of a white solid. The resulting suspension was kept for 15 h at room temperature and then filtered and vigorously rinsed with deionised water in order to remove the remaining ammonia and chloride ions. The amorphous nature of the precursor has been recently analysed and reported else where [17].

#### 2.1.2 TNT synthesis

The TNT synthesis was performed in an autoclave system using ~6 g of precursor in *ca.* 60 ml of NaOH 10 M aqueous solution. The samples were prepared at 200 ºC using an autoclave dwell time of 12 hours. After cooling, the suspension was filtrated and the solid was washed systematically: the solid was dispersed in ~100 mL of water and magnetically stirred for one hour. Then, the suspension was filtered and the pH of the filtrate measured for reference. This procedure was repeated until *pH*=9 (sample TNT-pH9) was achieved in the filtrate solution. The sample TNT-pH7 corresponds to a final filtrate with *pH*=7. These two solids were dried and stored. The wet-solid TNT-pH7 was also used to prepare a protonated sample (TNT-pH5) by dispersing and stirring it in a $HNO_3$ 0.1 M aqueous solution for 1 hour. Next, the solid was washed with water until a filtrate with *pH*=5 was obtained. This solid was then dried and stored. With this simple procedure, which consists only in controlling the filtrate *pH* (reference *pH*) through the above mentioned washing steps, the preparation of materials with different sodium/proton ratio was achieved.

**2.2 Adsorption experiments**

Before photocatalytic experiments, some adsorption studies were carried out using a dye-TNT aqueous suspensions (1g/L) under stirring for 1 hour in dark conditions. After centrifugation, the dyes' concentrations were estimated by measuring the absorbance at each dye chromophoric peak. In order to better understand the surface phenomena involved in the adsorption process different cationic and anionic dyes were used. The following dyes were tested: nafthol yellow S, methyl orange, rhodamine 6G and methylene blue.

**2.3 Photodegradation experiments**

All photodegradation experiments were conducted using a 250 ml refrigerated photoreactor [18]. A 450 W Hanovia medium-pressure mercury-vapour lamp was used as radiation source, the total irradiated energy being 40-48% in the ultraviolet range and 40-43% in the visible region of the electromagnetic spectrum.

Suspensions were prepared by adding 25 mg of powder to 125 ml of 10 ppm dye aqueous solution. Prior to irradiation, suspensions were stirred in darkness conditions for 60 min to ensure adsorption equilibrium. During irradiation, suspensions were sampled at regular intervals, centrifuged and analysed by UV-vis spectroscopy.

**2.3 Characterization**

X-ray powder diffraction was performed using a Philips X-ray diffractometer (PW 1730) with automatic data acquisition (APD Philips v3.6B), using Cu Kα radiation ($\lambda$=0.15406 nm) and working at 40 kV/30 mA. The diffraction patterns were collected in the range $2\theta = 7\text{-}60°$ with a 0.02° step size and an acquisition time of 2.0 s/step. The diffractometer was calibrated before each measurement. Optical characterization of the samples was carried out by UV-Vis diffuse reflectance using a Shimadzu UV-2450PC spectrometer. Diffuse reflectance spectra (DRS) were recorded in the wavelength range of 220-850 nm. Transmission electron microscopy (TEM) and high resolution transmission electron microscopy (HRTEM) were carried out using a JEOL 200CX microscope operating at 200 kV. A UV-vis spectrophotometer Jasco V560 was used for monitoring the absorption of the dye solutions. Zeta potential measurements were performed with a Zeta Sizer Nano Series (Malvern), using an aqueous 0.1 M HCl solution for *pH* adjustments.

## 3. Results and discussion

### 3.1. Identification and morphology

The XRD patterns of the TNT-pH5, TNT-pH7 and TNT-pH9 samples are presented in figure 1. For all the prepared materials, the XRD patterns are in agreement with the existence of a $Na_xH_{2-x}Ti_3O_7$ titanate layered structure, with $x$ decreasing with the reference *pH* (filtrate's *pH*). The diffraction peak identified at $2\theta \sim 10°$ is related with the interlayer distance while the peaks at 24.5°, 28.6° and 48.6° are characteristic of tri-titanate 1D nanomaterials [19]. As can be seen, when the reference *pH* decreases, the peak at ~10° shifts to higher $2\theta$ values, changing from 10.46 for the TNT-pH9 sample to 10.58 for TNT-pH7 and finally to 11.10 for the TNT-pH5 material. Moreover, following the above mentioned $2\theta$ shift to higher values, a broadening of this peak was also observed when the reference *pH* decreases. These structural features are related with different sodium contents in the samples [19,20], and are indicative of a $Na^+ \rightarrow H^+$ substitution process in the interlayer region. Moreover, the decrease of the $Na_2Ti_3O_7$ peak intensity (*e.g.* $2\theta = 25.7°$), and the simultaneous increase of the $H_2Ti_3O_7$ peak intensity (*e.g.* $2\theta = 48.5°$) also support the sodium/proton replacement in the crystalline structure referred to above.

The TNT samples were analysed by transmission electron microscopy and no changes in the elongated tubular morphology were seen for the three samples. In order to corroborate the occurrence of an adjustment in the crystalline structure due to the $Na^+ \rightarrow H^+$ replacement, the interlayer distance of each sample was directly measured from HRTEM images (figure 2) and compared with the data obtained through the XRD. The values obtained were 1.13±0.03 nm for the TNT-pH9, 0.91±0.05 nm for the TNT-pH7 and 0.70±0.02 nm for the TNT-pH5 sample. It has been reported in the literature [21] that the distance between layers varies from ~ 0.70 nm in the protonated TNT, to ~0.90 nm in the alkaline form. The results obtained are consistent with a sequential sodium/proton replacement as the number of washing steps increases. Note that the distance between layers obtained for the TNT-pH9 sample is greater than the value reported by Bavykin et al. [21]. This difference may be due to a higher sodium content in our sample. Furthermore, the result obtained for the TNT-pH5 sample agrees with a complete sodium replacement in this material.

**3.2. UV-vis photo-response**

Because of its relevance in photocatalytic processes, the photo-response of the prepared TNT materials was studied. Optical characterization of the samples was carried out by measuring their diffuse reflectance ($R$) spectra at room temperature. $R$ is related with the absorption Kubelka-Munk function, $F_{KM}$, by the relation $F_{KM}(R) = (1-R)^2/2R$, which is proportional to the absorption coefficient [22]. The $F_{KM}$ absorption spectra are shown in figure 3a. As can be seen, a significant red shift in the optical absorption band edge can be observed for the protonated sample (TNT-pH5) compared with the TNT-pH9 and TNT-pH7 samples. The optical band gap energies of the TNT samples were calculated by plotting the function $f_{KM} = (F_{KM} h\upsilon)^{0.5}$ vs. energy (Tauc plot), where $h$ stands for the Planck constant and $\upsilon$ for the frequency. The linear part of the curve was extrapolated to $f_{KM} = 0$ to get the indirect band gap energy of each material (fig. 3b). The estimated $E_g$ values were 3.30±0.06 eV for TNT-pH7 and 3.27±0.03 eV for the TNT-pH9 material. These values are in agreement with previously published works [23,24]. Note however that the $E_g$ value obtained for the TNT-pH5 sample, 2.81±0.02 eV, is much lower than the $E_g$ values of the other two samples, bringing its absorption edge to the near-visible region, in agreement with the above mentioned absorption spectra. This fact opens the possibility of using TNT-pH5 as a photoactive material in a broader range of the electromagnetic spectrum, and may contribute to the enhanced photocatalytic activity of this sample, as it will be shown later.

**3.3 Point of zero charge calculation**

In view of the effect of surface charge on the adsorption processes, the point of zero charge (p.z.c.) of the TNT samples was measured. When prepared by alkaline hydrothermal synthesis in aqueous media, TNT are negatively charged and are expected to attract cationic species due to the presence of cation binding sites, which have been formed during the synthesis by $Na^+$ incorporation [25]. Through sodium/proton exchange, the protonated nanotubular titanates tend to develop a negative zeta potential due to the dissociation of titanic acid according to the following reaction:

$$H_2Ti_3O_7 \rightleftarrows H^+ + HTi_3O_7^- \qquad (1)$$

This phenomenon can have an effect on the ability of the charged molecules to be adsorbed at the TNT surface [21].

The *pH* value at which the TNT surface carries no net charge (p.z.c.) was evaluated for all the prepared materials, as shown in figure 4. The p.z.c. values obtained for the TNT-pH9 and TNT-pH7 samples were 3.2 and 3.1, respectively. These results demonstrate that these materials are negatively charged over most of the *pH* range (higher than *ca.* 3) and are in agreement with some studies reporting a p.z.c. of *ca.* 3 for titanate nanotubular structures [21,25]. Analysis of figure 4 allows to conclude that TNT-pH9 and TNT-pH7 samples display a positive and similar zeta potential trend for *pH* values less than their p.z.c. values. Note however that for *pH* higher than the corresponding p.z.c. values, a distinct behaviour was observed for those two samples. This result should be interpreted considering that TNT-pH9 and TNT-pH7 samples have some ion-exchange ability. For instance for the TNT-pH9 sample a large amount of $Na^+$ remains in the structure and can easily be exchanged with $H^+$ from the solution. For the protonated sample (TNT-pH5), a p.z.c. of 3.9 was estimated. The shift observed for the p.z.c. value of this sample, compared to the other two higher sodium contents samples, can be directly related with its acidic characteristics (see equation 1).

### 3.4. Dye adsorption

Since the effect of surface adsorption phenomena on the photocatalytic processes is important, a preliminary study of the adsorption affinity of several cationic and anionic dyes for the TNT surfaces was carried out. For the purposes of comparing adsorptions, and based on the TEM results, the surface area was assumed to be constant for the three TNT materials. As expected on the basis of the estimated p.z.c. values, the anionic naftol yellow S and methyl orange dyes did not exhibit any significant adsorption affinity for the three TNT samples. In contrast, the cationic methylene blue and rhodamine 6G dyes were strongly adsorbed by all the TNT samples. These results confirms that the surfaces of the three TNT samples are negatively charged when in contact with the aqueous cationic dye solutions and corroborates the fact that the interactions involved in the adsorption phenomena are predominantly electrostatic [21].

### 3.4.1. R6G adsorption

The R6G adsorption profiles of the prepared samples are shown in figure 5. As can be seen, this cationic dye was adsorbed at all the TNT sample surfaces, yet at considerable different extents. The material with the best R6G adsorption ability was TNT-pH9. Under the experimental conditions used, 62.83 % of the initial R6G present in solution was adsorbed at the TNT-pH9 surface. Using identical conditions, adsorption by the TNT-pH7 and TNTpH5 samples drops to 30.59% and to 9.37%, respectively. The analysis of these results should consider the influence of two factors in the dye adsorption process: i) the surface ionic charge and ii) the possibility of interlayer R6G intercalation The difference in the surface ionic charge, of the three materials, does not by itself explain the adsorption results since the numerical values for TNT-pH9 and TNT-pH7 p.z.c. are very close (3.1 and 3.2, respectively) and the corresponding R6G adsorption values are very different (62.83% and 30.59%, respectively). On the other hand, the occurrence of intercalation phenomena between the $TiO_6$ layers may be possible for TNT samples with higher reference *pH* since they have sodium ions available to be exchanged. However, the dye molecule dimensions must be considered in order to validate the possibility of a R6G intercalation process. According with values reported in literature [26], a R6G molecule presents a plane with dimensions $1.4 \times 1.4$ nm$^2$ and a thickness of ~0.85 nm. The latter value is lower than the interlayer distance measured for samples TNT-pH9 (1.13 nm) and TNT-pH7 (0.91 nm). Consequently, if the R6G molecule plane is parallel to the $TiO_6$ layers, the dye molecule may be incorporated into the interlayer space of those two types of nanostructures, the $TiO_6$/R6G/$TiO_6$ intercalation process being easier for sample TNT-pH9 than for sample TNT-pH7. Concerning TNT-pH5 sample, its interlayer distance is lower than the R6G molecule thickness, and thus the intercalation process can not play a key role in the adsorption process of R6G. Nevertheless, it should be noted that some degree of $TiO_6$/R6G/$TiO_6$ intercalation can not be ruled out in the TNT-pH5 sample since adjustments in the TNT lamellar structure are possible and have been reported during intercalation processes of outsized organic molecules [27]. Therefore, not only the surface ionic charge differences but also the R6G intercalation hypothesis seems to support the different R6G adsorption abilities of the TNT samples prepared.

### 3.5. R6G photocatalytic degradation

The effect of the sodium/proton content on the TNTs catalytic activity was studied in the R6G dye photodegradation process, by using UV-vis spectroscopy. The typical absorption spectra of R6G is characterized by three main bands, one in the visible region, $\lambda_{max}$ = 528 nm, which is related to the chromophoric group (the dye colour arises from the aromatic rings connected by azo groups) and the other two in the UV region, at $\lambda_{max}$ = 247 nm and $\lambda_{max}$ = 275 nm, related with the absorption of benzene and naphthalene-like structures in the molecule, respectively. In a R6G photodegradation process, the decrease of the 528 nm band intensity is used to quantify the dye's decolorization degree, while the decrease of the naphthalene ring absorption band intensity, at 275 nm, can be used to infer its degradation degree [28]. The absorbance peak at 247 nm is attributed to benzene-like structures arising from R6G degradation. Figure 6 shows the absorption spectra obtained in 90 minutes of irradiation. A negative time (-60 min) was used for convenience to show the dye adsorption phenomenon before irradiation. Distinct dye absorption spectral profiles were observed during photolysis and when using the three different TNT samples as photocatalysts, which suggest different chemical pathways for the R6G photocatalytic degradation process.

On the basis of the variation of the 528 nm absorption peak intensity, the dye photodecolorization profiles were calculated and are shown in Figure 7. As can be seen, all the TNT samples exhibit photocatalytic activity. A 97.6 % decrease of the initial dye concentration was achieved after 90 min of irradiation using the TNT-pH5 sample. A similar R6G concentration decrease (~93 %) was obtained when using either the TNT-pH7 or the TNT-pH9 sample as photocatalytic materials. Note that only a 37 % reduction of the initial R6G concentration was reached without catalyst (photolysis). It is interesting to point out that the best R6G photocatalytic performance was achieved by the material with the lowest dye adsorption capability (TNT-pH5).

Beyond the absorption band associated with the chromophoric groups (528 nm), it is also important to analyze the variation of the intensities of the absorption bands associated with the naphthalene- (275 nm) and benzene-*like* structures (247 nm), since they are related with hazard species that should be degraded. Figure 8 shows the relative intensities ($I/I_0$) of those two absorption bands plotted as a function of irradiation time for each sample. As can be seen, by using the TNT-pH5 catalyst a slight increase of both band intensities was observed for the first 5 minutes of irradiation. Afterwards, their $I/I_0$ values continuously decrease. Note, however, that after 90 minutes of irradiation the existence of naphtalene and bezene-*like* structures can still be inferred, though at much lower quantities than initially. These results show that not only was the dye's photodecolorization achieved, as shown before in figure 7, but also a substantial effective degradation of the R6G (figure 8) when using the TNT-pH5 sample as photocatalys. A different behaviour was found for the degradation processes when using TNT-pH7 and TNT-pH9 samples. As can be seen in figure 8, the relative intensities of the bands at 275 nm and 246 nm increase with irradiation time. This means that, despite the photodecolorarization effectiveness of the R6G using these two materials as photocatalysts, the overall process leads to the formation of new aromatic secondary products in substantial amounts. Note that this behaviour is much more pronounced when using TNT-pH7 than when using TNT-pH9. Work is in progress to clarify the R6G degradation reaction mechanisms related with these different photocatalytic behaviours.

## 4. Conclusions

Homogenous and stable titanate nanotubular materials with different sodium contents were prepared using a hydrothermal chemical route which avoids the use of crystalline $TiO_2$ as precursor material. Nanomaterials with different $Na^+/H^+$ ratio were obtained using a swift control of the washing steps. XRD and HRTEM analyses showed that the $Na^+ \rightarrow H^+$ replacement leads to a shortening TNT interlayer distance. Depending on the sodium/proton ratio different optical behaviours were observed. A band gap energy of $3.27\pm0.03$ eV was estimated for the material with higher sodium content. This value was red shifted to $2.81\pm0.02$ eV for the protonated material, bringing its absorption edge into the near visible region. A p.z.c. value of 3.9 was estimated for the protonated sample while p.z.c. values of 3.2 and 3.1 were estimated for samples TNT-pH9 and TNT-pH7, respectively. The adsorption ability and photocatalytic activity were studied using the R6G organic dye. The best photocatalytic activity was achieved by the protonated material (TNT-pH5); the best adsorption ability was, however, exhibited by the higher sodium content sample (TNT-pH9). Finally, analysis of the dye solution absorption spectra seems to provide evidence for the existence of distinct chemical pathways for the photocatalytic R6G degradation process.

**Acknowledgments**

This work was supported by Fundação para a Ciência e Tecnologia (PTDC/CTM-NAN/113021/2009). The authors thank P.I. Teixeira for a critical reading of the manuscript.

# Figure captions

**Figure 1** - XRD patterns of the TNT-pH9, TNT-pH7 and TNT-pH5 materials prepared at 200ºC for 12 hours. Symbols: (•) - $Na_2Ti_3O_7$, JCPDS-ICDD file No. 31-1329, (∗) - $H_2Ti_3O_7$, JCPDS-ICDD file No. 41-192. The inset shows the $2\theta$ range around 10º, where the peaks related with the TNT interlayer distance are located.

**Figure 2** - HRTEM images of the TNT samples prepared. a) TNT-pH9; b) TNT-pH7; c) TNT-pH5.

**Figure 3** - a) Absorption spectra and b) Tauc plots of the TNT samples.

**Figure 4** - zeta potential *versus* reference pH for the TNT materials prepared.

**Figure 5** - R6G adsorption ability for the TNT-pH9, TNT-pH7 and TNT-pH5 surfaces.

**Figure 6** - Absorption spectra of the R6G solution during photo-irradiation: a) photolysis, b) TNT-pH5, c) TNT-pH7 and d) TNT-pH9 as photocatalyst.

**Figure 7** - Photocatalytic decolourization of a 10 ppm R6G aqueous solution (125 ml) using 25 mg of catalyst.

**Figure 8** - Relative intensities of the R6G's absorption bands at $\lambda = 275$ nm and $\lambda = 275$ nm as a function of irradiation time. Symbols: --□-- photolysis; --■-- using TNT-pH5 as photocatalyst; --▲-- using TNT-pH7 as photocatalyst; --▼-- using TNT-pH9 as photocatalyst.

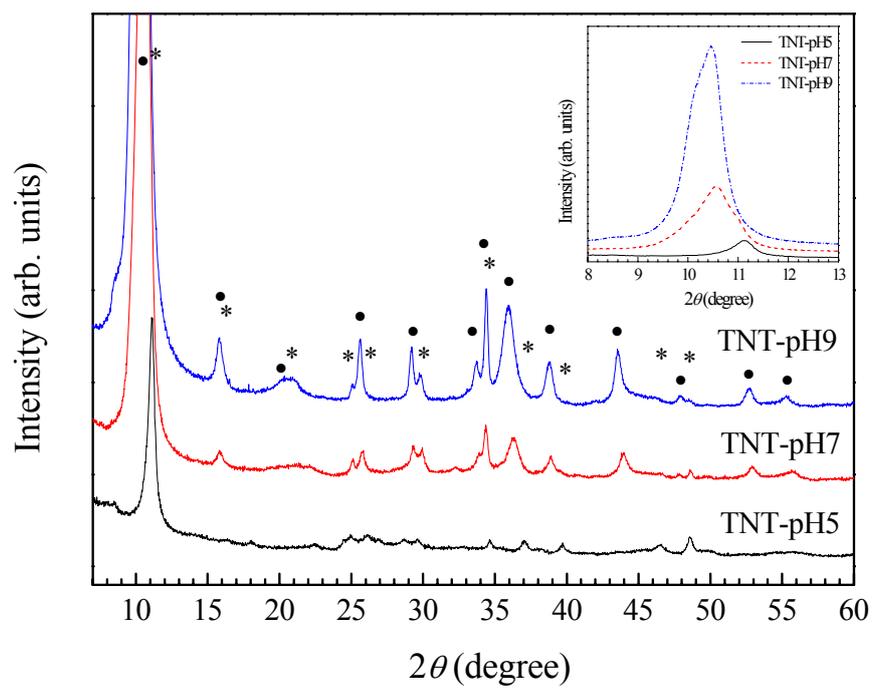

Figure 1

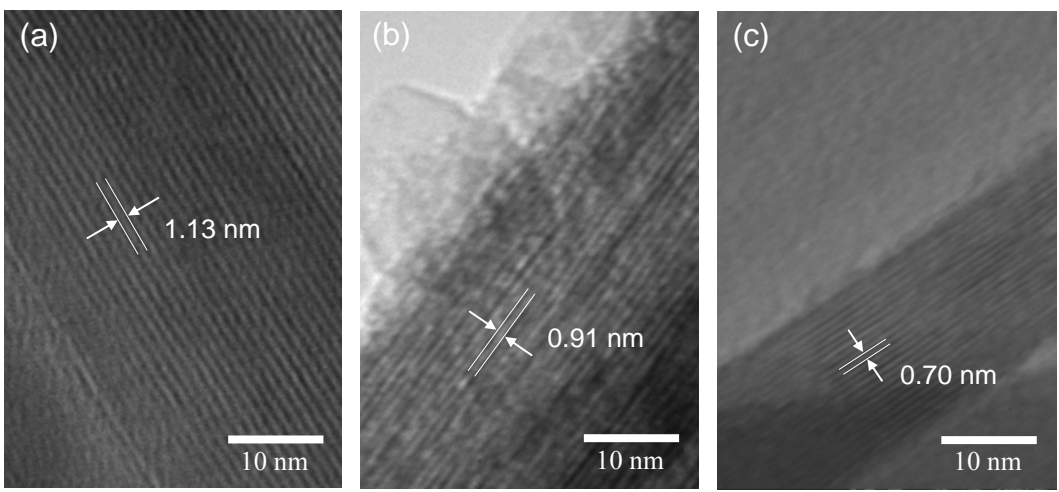

Figure 2

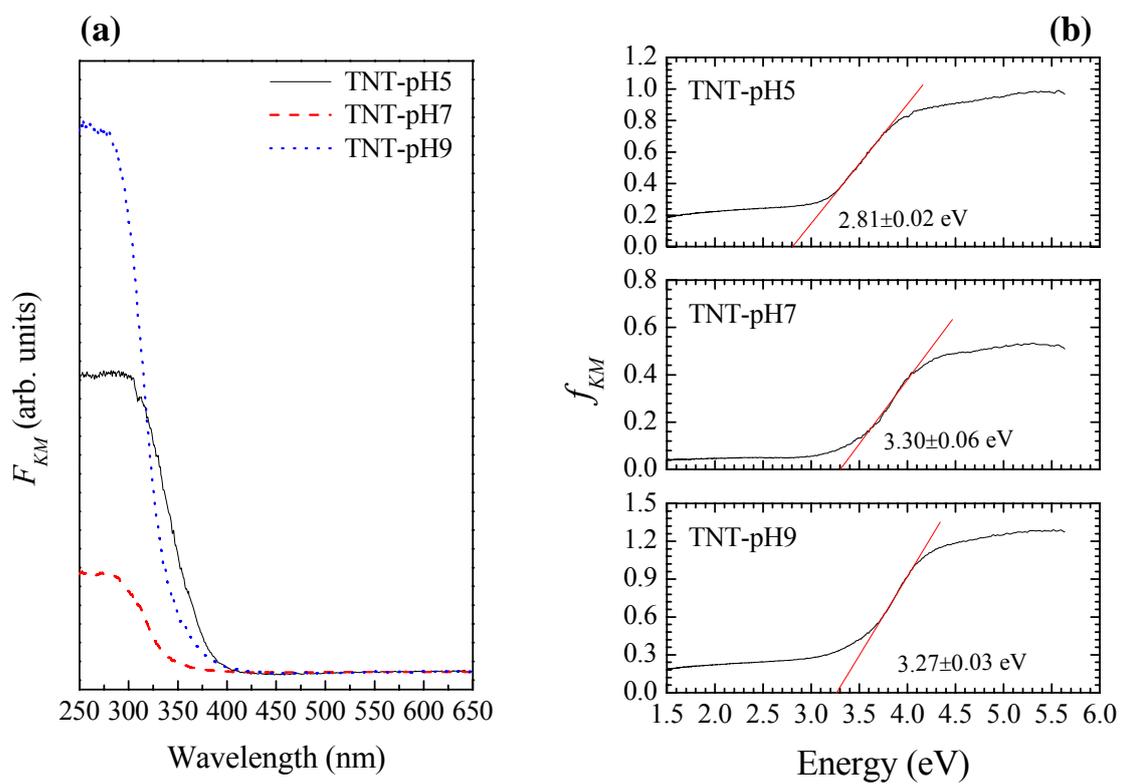

Figure 3

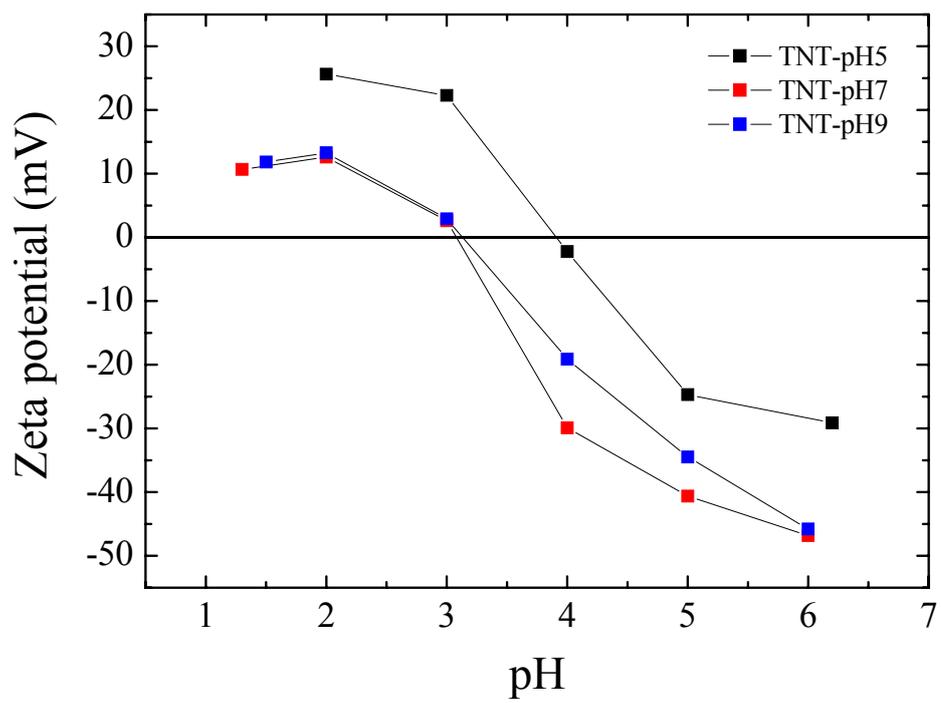

Figure 4

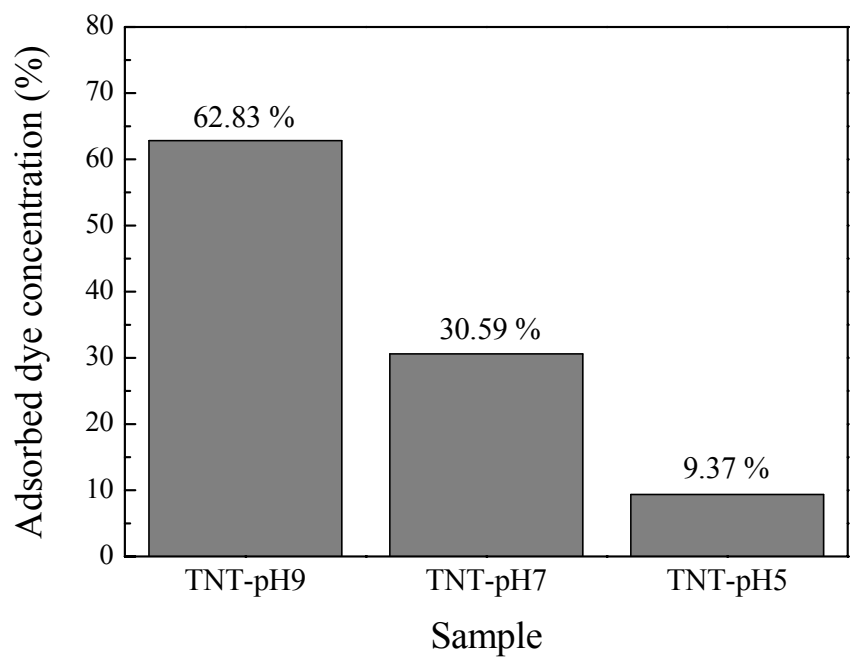

Figure 5

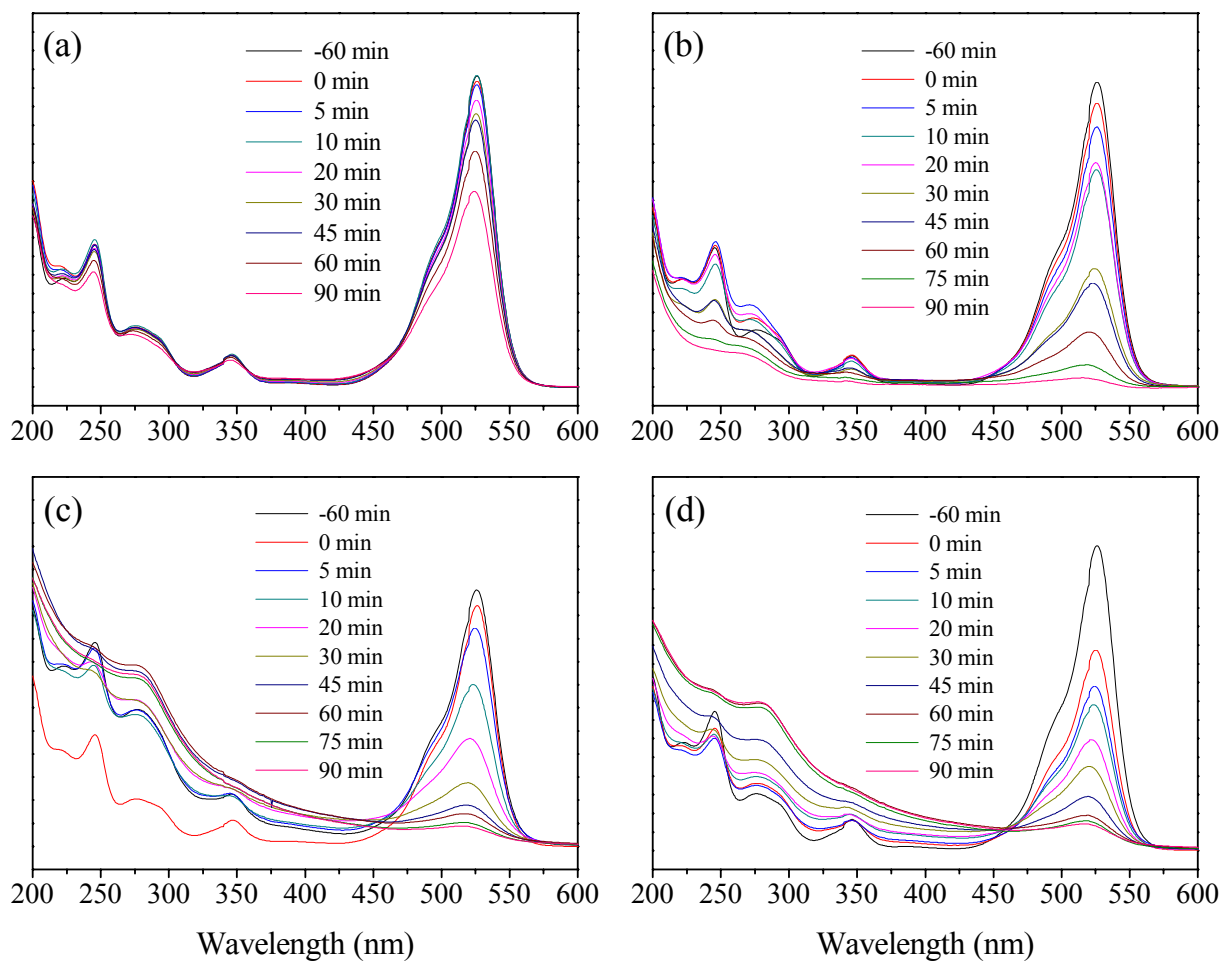

Figure 6

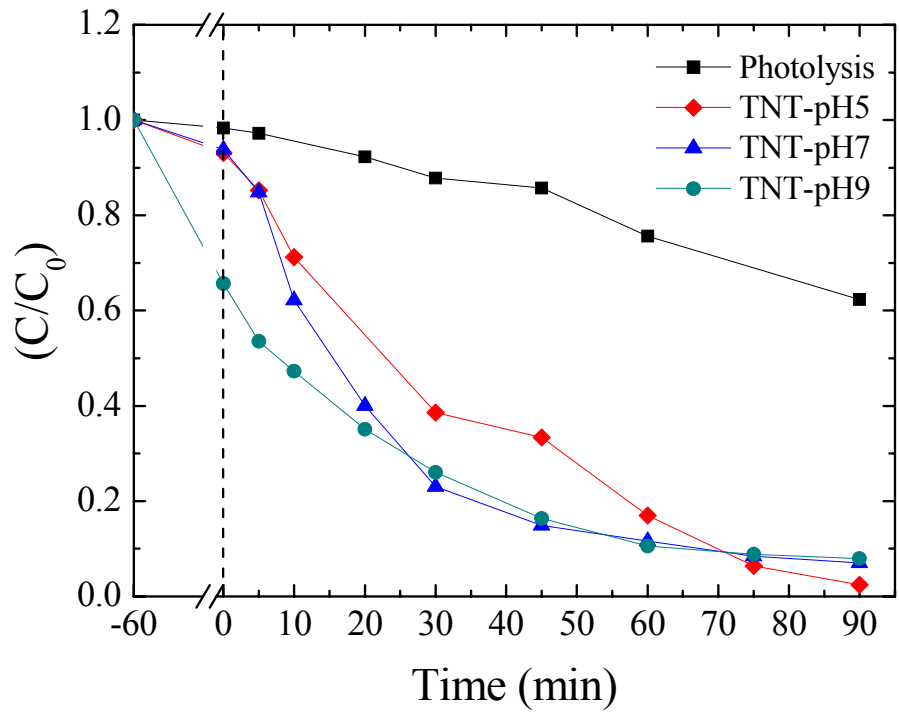

Figure 7

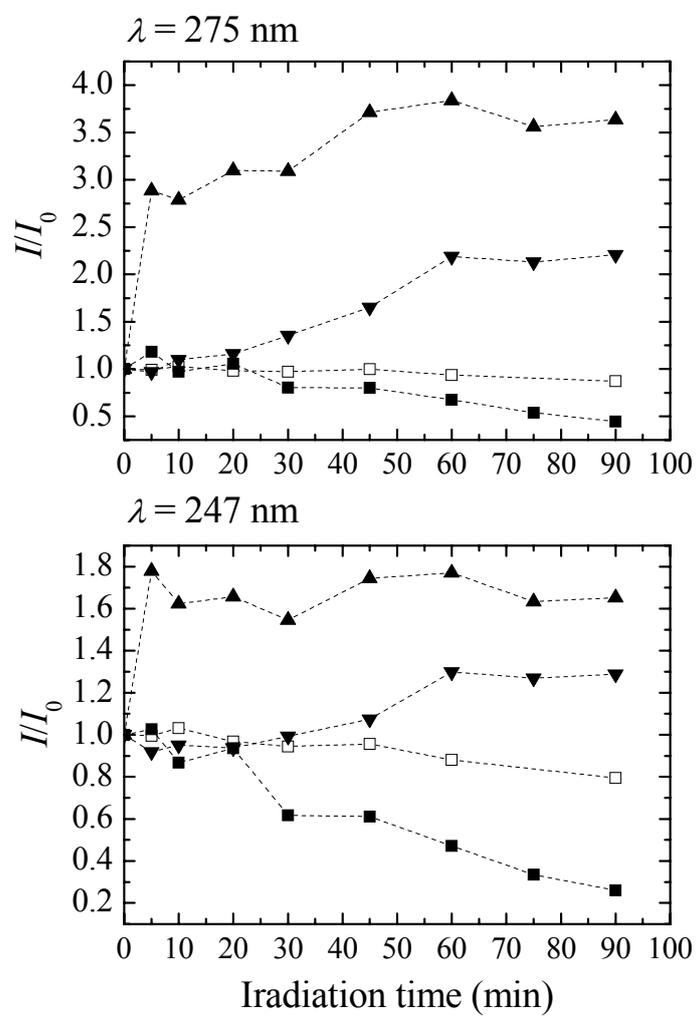

Figure 8